\documentclass[11pt,twoside]{article}
\usepackage{baaa2011}
\usepackage{graphicx}
\usepackage{subfigure}
\usepackage{psfrag}
\usepackage{amssymb}
\usepackage[spanish,activeacute,english]{babel}
\usepackage[latin1]{inputenc}
\usepackage[T1]{fontenc} 
\usepackage{ae,aecompl} 
\usepackage{latexsym}
\usepackage{verbatim}
\usepackage{amsmath}
\usepackage{stmaryrd}
\usepackage{amsfonts}
\usepackage{amssymb}
\usepackage{wasysym}
\usepackage[colorlinks=true,dvips]{hyperref}

\begin{document}
\myselectspanish 
\vskip 1.0cm
\markboth{C. Moni Bidin et al.}%
{Dark matter in the solar neighborhood}

\pagestyle{myheadings}
\vspace*{0.5cm}
\noindent PRESENTACI\'ON ORAL
\vskip 0.3cm
\title{No evidence of dark matter in the solar neighborhood}

\author{C. Moni Bidin$^{1}$, G. Carraro$^{2,3}$, R.A. M\'endez$^{4}$, R. Smith$^{1}$}

\affil{%
  (1) Universidad de Concepci\'on, Chile \\
  (2) ESO - Chile \\
  (3) Universit\'a di Padova, Italy \\
  (4) Universidad de Chile, Chile \\
}

\begin{abstract}
We measured the surface mass density of the Galactic disk at the solar position, up to 4~kpc from the plane,
by means of the kinematics of $\sim$400 thick disk stars. The results match the expectations for the visible
mass only, and no dark matter is detected in the volume under analysis. The current models of dark matter halo
are excluded with a significance higher than 5$\sigma$, unless a highly prolate halo is assumed, very atypical
in cold dark matter simulations. The resulting lack of dark matter at the solar position challenges the
current models.
\end{abstract}

\begin{resumen}
La densidad superficial de masa del disco Galactico a la posici\'on solar fue medida, hasta 4~kpc del plano,
a trav\'es de la cinem\'atica de $\sim$400 estrellas del disco grueso. Los resultados coinciden con las
expectaciones para la sola materia visible, y no se detecta materia obscura en el volumen analizado.
Los modelos actuales de halo de materia obscura son excluidos a nivel mayor de 5$\sigma$, a menos que se
asuma un halo fuertemente prolado, muy at\'ipico en las simulaciones de materia obscura fr\'ia. La consiguiente
falta de materia obscura a la posici\'on solar desaf\'ia los modelos actuales.
\end{resumen}

\section{Introduction}
Measuring the matter density of the Galactic disk by means of the spatial distribution and kinematics of
its stars is an old art, dating nearly a century (Kapteyn 1922, Oort 1932). The comparison of the results
with the expected amount of visible matter provides an estimate of the dark matter (DM) density in the
analyzed volume. So far, all but few estimates converged to the conclusion that ``there is no evidence
for a significant amount of DM in the Galactic disk" (e.g., Kuijken \& Gilmore 1989; Holmberg \& Flynn 2004).
Apart from this very general statement, whose interpretation is not even unique (see Garbari et al. 2011; for
a discussion), only little progress has been made on constraining the fundamental properties of the DM halo,
such as its flattening and local density. This is very unfortunate, because the shape of the dark halo bears
information about the nature of the DM itself (Olling \& Merrifield 2000). Moreover, the results of the
experiments for direct detection of DM are degenerate between the unknown interaction cross-section of the
searched particles and their local density. Therefore, the local DM density of the Standard Halo Model
(SHM, $\rho_{\odot,DM}$=8$\cdot 10^{-3}$~M$_\odot$~pc$^{-3}$, Jungman et al. 1996) have so-far been assumed
in their interpretation. This density, however, is only a mean value compatible with indirect evidences such
as the Milky Way rotation curve.

The strongest limitations on the measurements of the Galactic dynamical mass come from the great
observational effort required to derive the spatial distribution of a stellar population and the variation
of its tree-dimentional kinematics. For this reason, approximations have always been introduced in the
calculations, whose validity, often questioned (e.g., Siebert et al. 2008; Garbari et al. 2011), decreases
with distance from the Galactic plane. As a consequence, all previous investigations have been limited to
$\pm$1.1~kpc from the plane, but the amount of DM in this volume is small compared to the observational
errors, and firm conclusions are prevented.

\section{Results}

We estimated the dynamical mass at the solar Galactocentric position between $Z$=1.5 and 4~kpc from the
plane, as inferred by the variation of the kinematics of the Galactic thick disk with $Z$. This was
measured by Moni Bidin et al. (2012), who analyzed a sample of $\sim$400 red giants with 2MASS photometry
(Skrutsie et al. 2006), SPM3 proper motion (Girard et al. 2004), and radial velocity (Moni Bidin 2009).
Their kinematical results were inserted into Equation~2 of Moni Bidin et al. (2010), that was obtained
inserting the Jeans equations into the Poisson equation and integrating. This equation is exact within
the limits of validity of simple symmetry requirements, plus a set of additional assumptions, namely:
$i$) steady state; $ii$) radial and vertical exponential decay of the density; $iii$) flat rotation
curve; $iv$) no disk flare; $v$) constancy of the radial scale lenght with distance from the plane; $vi$)
radial exponential decay of the dispersions, with the same scale length as the mass density. The three
required input parameters (solar Galactocentric distance, thick disk scale height and length) were
defined by the average of about 20 literature estimates (Moni Bidin et al. 2010).

\begin{figure}[!t]
\centering
\includegraphics[width=.45\textwidth]{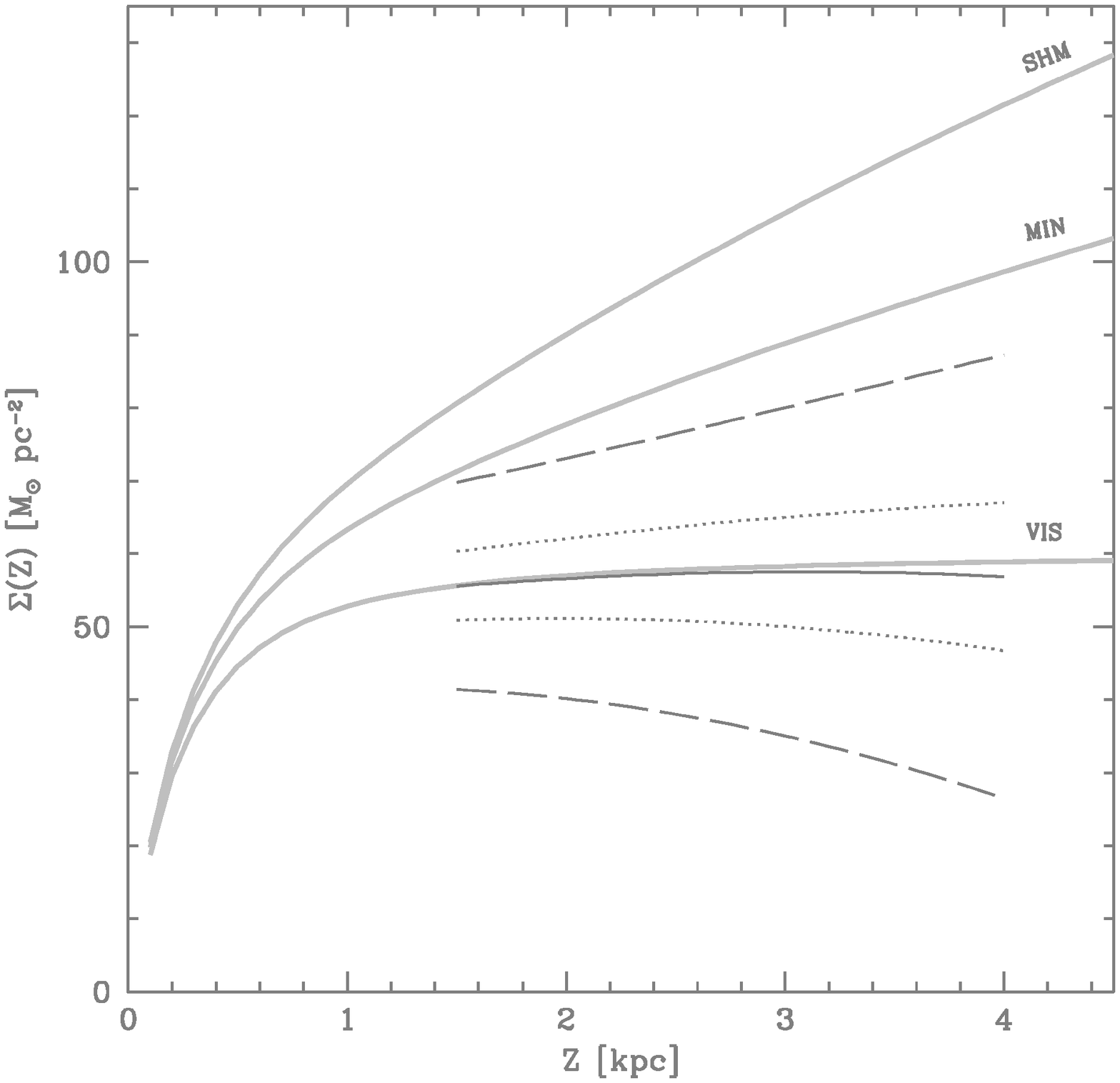}~\hfill
\includegraphics[width=.45\textwidth]{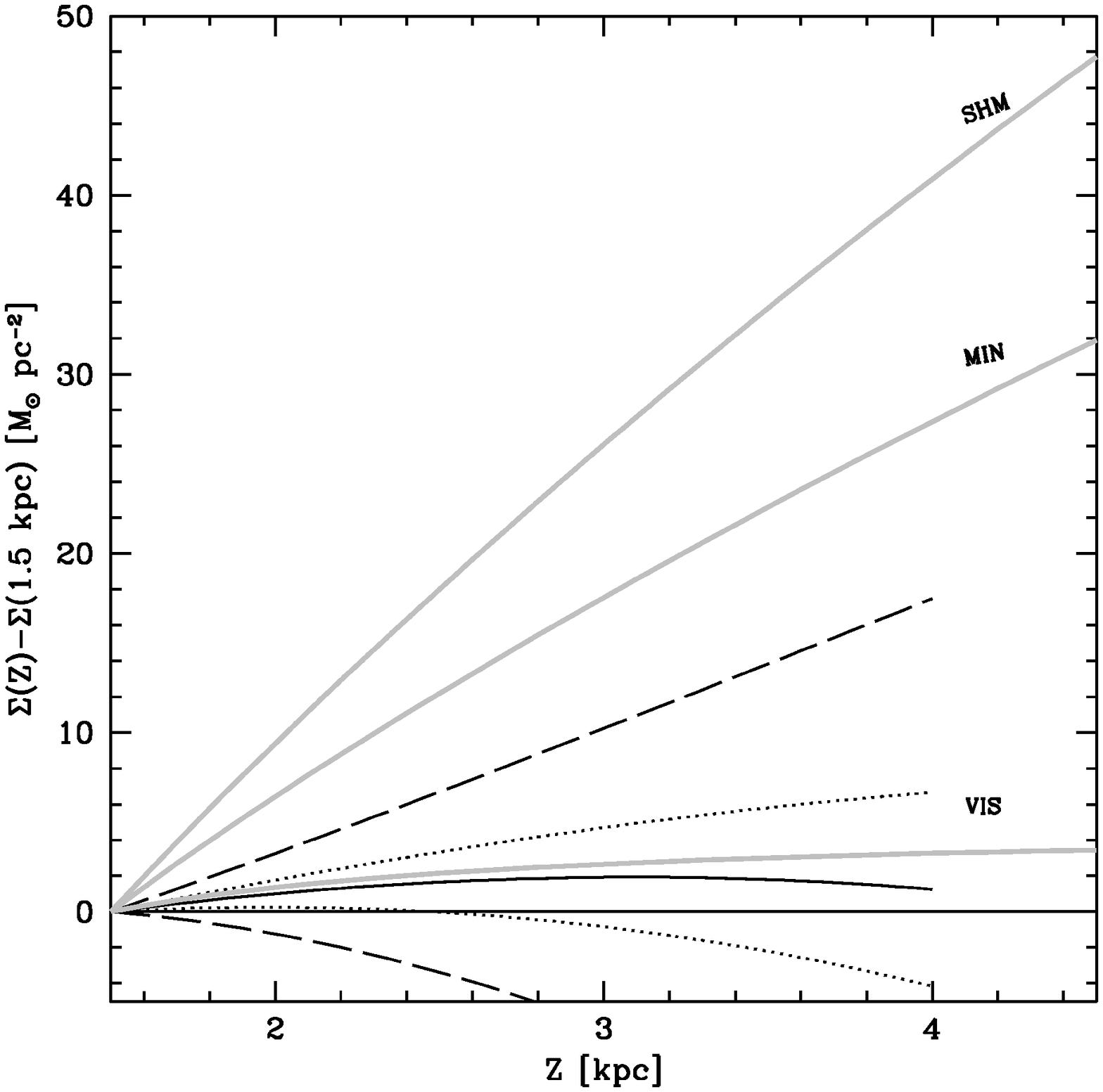}
\caption{Calculated absolute (left panel) and incremental (right panel) surface mass density, as a
function of distance from the Galactic plane. The dotted and dashed curves indicate the 1$\sigma$ and
3$\sigma$ strip, respectively. The expectations of the known visible mass (VIS), and of two models
(SHM and MIN) comprising the visible mass plus the DM halo described in the text.}
  \label{f_res}
\end{figure}

The results of our calculations are shown in the left panel of Figure~\ref{f_res}, where they are
compared to the known amount of visible matter, as estimated by Moni Bidin et al. (2010). The
expectations of two spherical Navarro et al. (1997) models for the DM halo, with the local density equal
to the SHM (labeled SHM), and to the minimum density extrapolated by the Galactic rotation curve
($\rho_{\odot,DM}$=5$\cdot 10^{-3}$~M$_\odot$~pc$^{-3}$, MIN model; Weber \& de Boer 2010), are also
overplotted. The derived surface density $\Sigma (Z)$ well matches the expectations for visible mass
alone, and no DM is detected in the volume under analysis. From the derived curve, a local DM density of
0$\pm$1$\cdot 10^{-3}$~M$_\odot$~pc$^{-3}$ is derived. The SHM model is therefore excluded at the
8$\sigma$ level, and even the model with the minimum density (MIN) is 5$\sigma$ more massive than the
detected dynamical mass. Interestingly, very similar results are obtained if the kinematical results of
Casetti-Dinescu et al. (2011) are assumed in the calculations. In fact, the curve thus derived matches
the expectations for the visible mass only, although with much lower significance because of the larger
errors. Identical conclusions are drawn if the integration of the Poisson equation is performed in the
interval 1.5--4~kpc instead of 0--4~kpc (right panel of Figure~\ref{f_res}). This calculation is more
reliable, because the kinematics is not extrapolated at $Z\leq$1.5~kpc, where it was not measured.
Moreover, the uncertainty on the quantity of visible mass is also limited, because this range is above
the thin layer of disk interstellar medium, and it encloses only the tail of the Galactic stellar disk
distribution.

It can be shown, by means of extensive calculations, that altering one of the hypothesis or the value of
one of the three parameters cannot solve the problem of the missing DM in the volume under analysis. The
details of this analysis will be given in a forthcoming paper (Moni Bidin et al. 2012). The derived
solution can be forced to match the expectations of the DM halo models only under an exotic combination
of unlikely hypothesis as, for example, a very thin thick disk (scale height 0.7~kpc) either very
extended in the radial direction (scale length 4.6~kpc) or strongly flared at the solar position. On the
contrary, the expected visible mass matches the observations without any effort, by use of the most
probable assumptions. The models can reproduce the observed curve assuming a highly prolate DM halo,
because the local quantity of DM is inversely proportional to the flattening $q$ of the spheroidal
distribution. The requirement that the least massive model (MIN) agrees with the observations within
2$\sigma$ returns the constraint $q\geq 2$. Nevertheless, current cold DM simulations have problems in
reproducing such strongly prolate structures (e.g., Dubinski \& Carlberg 1991), and this solution would
therefore require a revision of the models.

\section{Conclusions}

The observations point to a noticeable lack of DM at the solar Galactocentric position. It is easy to
see that the presence of a classical DM halo as those shown in Figure~\ref{f_res} would have been
unequivocally detected with our method even if, as suggested by Moni Bidin et al. (2010), it could
generate only a small variation of the potential. In fact, inserting the derivative of the potential of
any DM halo model into the integrated Poisson equation, no mismatch arises between the resulting
dynamical mass and the quantity of DM mass enclosed between $\pm Z$. Moreover,
S\'anchez-Salcedo et al. (2011) showed that the presence of a DM halo affects the disk kinematics
noticeably (compare their Figure~1 and~2), and the difference is much higher than the observational
errors of Moni Bidin et al. (2012). In conclusion, the interpretation of the observed lack of DM at the
solar position is not straightforward: DM is required to sustain the flat Galactic rotation curve, but
the observations point to a distribution very different to what today accepted. In particular, while
numerous experiments seek to directly detect the elusive DM particles our results suggest that their
density may be negligible in the solar neighborhood.

\agradecimientos
C.M.B. and R.A.M. acknowledge support from the Chilean Centro de Astrof\'isica FONDAP No. 15010003,
and the Chilean Centro de Excelencia en Astrof\'isica y Tecnolog\'ias Afines (CATA) BASAL PFB/06.
                                                                                
\begin{referencias}
\reference Casetti-Dinescu, D.I., Girard, T.M., Korchagin, V.I., \& van Altena, W.F. 2011, ApJ, 728, 7
\reference Dubinski, J., \& Carlberg, R.G. 1991, ApJ, 378, 496
\reference Garbari, S., Read, J.I., \& Lake, G. 2011, MNRAS, in press, arXiv:1105.6339
\reference Girard, T.M., Dinescu, D.I., van Altena, W.F., et al. 2004, AJ, 127, 3060
\reference Holmberg, J., \& Flynn, C. 2004, MNRAS, 352, 440
\reference Jungman, G., Kamionkowski, M., \& Griest, K. 1996, Physics Reports, 267, 195
\reference Kapteyn, J.C. 1922, ApJ, 588, 823
\reference Kuijken, K., \&  Gilmore, G. 1989, MNRAS, 239, 605
\reference Moni Bidin, C. 2009, PhD Thesis, Universidad de Chile, 1
\reference Moni Bidin, C., Carraro, G., M\'endez, R.A., \& van Altena, W.F. 2010, ApJ, 724, L122
\reference Moni Bidin, C., Carraro, G., \& M\'endez, R.A. 2012, ApJ, 747, 101
\reference Navarro, J.F., Frenk, C.S., \& White, S.D.M. 1997, ApJ, 490, 493
\reference Olling, R.P., \& Merrifield, M.R. 2000, MNRAS, 311, 361
\reference Oort, J.H. 1932, BAN, 6, 249
\reference S\'anchez-Salcedo, F.J., Flynn, C., \& Hidalgo-G\'amez, A.M. 2011, ApJ, 731, L35
\reference Siebert, A., Bienaym\'e, O., Binney, J., et al. 2008, MNRAS, 391, 793
\reference Skrutskie, M.F., Cutri, R.M., Stiening, R., et al. 2006, AJ, 131, 1163
\reference Weber, M., \& de Boer, W. 2010, A\&A, 509, A25
\end{referencias}

\end{document}